\documentstyle[twocolumn,prl,aps,epsf]{revtex}

\begin{document}
\draft


\title{Influence of uniaxial stress on the lamellar spacing of eutectics}

\author{Jens Kappey$^*$, Klaus Kassner$^*$ and Chaouqi Misbah$^{**}$}

\address{
$^*$ Institut f\"ur Theoretische Physik, Otto-von-Guericke-Universit\"at \\
Magdeburg, Postfach 4120, D-39016 Magdeburg, Germany
}

\address{
$^{**}$ Laboratoire de Spectrom\`etrie Physique, Universit\`e Joseph Fourier (CNRS),\\
Grenoble I - B.P. 87, 38402 Saint-Martin d' H\`eres Cedex, France
}

\maketitle

\begin{abstract}
Directional solidification of lamellar eutectic structures
submitted to uniaxial stress is investigated.
In the spirit of an  approximation first used by
 Jackson and Hunt, we calculate 
the stress tensor for a two-dimensional crystal with triangular surface,
using a Fourier expansion of the  Airy function.
The effect of the resulting change in chemical potential is
introduced into the standard model for directional solidification.
This calculation is motivated by an observation, made recently
[I. Cantat, K. Kassner, C. Misbah, and H. M\"uller-Krumbhaar,
Phys. Rev. E, in press], 
that the thermal gradient produces similar effects as a strong 
gravitational field in the case of dilute-alloy solidification.  
Therefore, the coupling between the Grinfeld and the 
Mullins-Sekerka instabilities becomes strong,
as the critical wavelength of the former instability gets reduced
to a value close to that of the latter.
Analogously, in the case of eutectics, 
the characteristic length scale of the Grinfeld instability should 
be reduced to a size not extremely far from typical lamellar spacings.
Following Jackson and Hunt, we assume the selected 
wavelength to be determined by the minimum undercooling criterion
and compute its shift due to the external stress.
In addition, we find that in general the volume fraction of the two solid 
phases is changed by uniaxial stress.
Implications for experiments on eutectics are discussed. 

\end{abstract}



\narrowtext

\section{INTRODUCTION}
\begin{figure}
\narrowtext
\epsfxsize=3in
\epsfbox{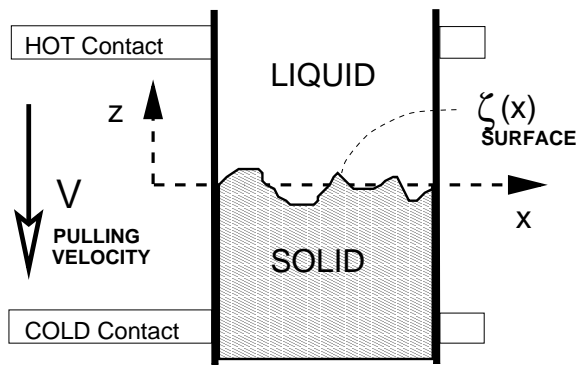}
\caption{
\noindent
Schematic setup of a directional solidification experiment. A container with the melt 
in it is pushed through a thermal gradiend G with a velocity V.
} 
\vskip 0.1in
\label{fig1} 
\end{figure}
A non-hydrostatically strained solid in contact
with its melt or vapor can partially relieve its elastic energy by
producing an undulated interface. This is the cause of a morphological 
instability  giving rise to the evolution of grooves with 
a definite spacing under uniaxial stress and, possibly, island formation, 
if the stress is biaxial. The instability
was first predicted by Asaro and Tiller\cite{Asaro72}.
Experimentally, it has been observed and studied by Torii and 
Balibar\cite{Torii92}.
Since the independent rediscovery of the
instability by Grinfeld\cite{Grinfeld93},
it has often been referred to as the Grinfeld or 
Asaro-Tiller-Grinfeld instability (ATG). Important contributions
leading to a broad interest in the instability are due to 
Nozi\`eres \cite{Nozieres92,Nozieres93}.

In directional solidification (Fig.\ref{fig1}), 
 it is known that the moving front undergoes, depending on the growth velocity, 
 another morphological instability, named after Mullins and Sekerka\cite{Mullins-Sekerka64} (MS),
 where the interface develops a cellular structure. Cantat {\it et al.}
\cite{Durand96,Cantat_et_al98} investigated the coupling between these 
two instabilities for dilute alloys.

\begin{figure}
\narrowtext
\epsfxsize=3in
\epsfbox{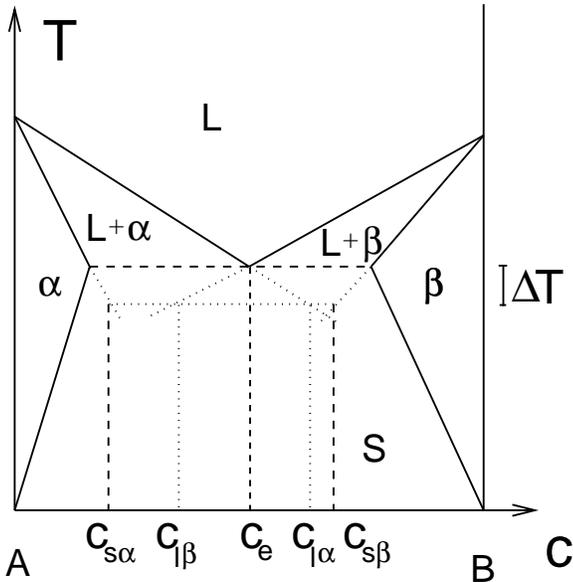}
\caption{
\noindent
 Generic phase diagram of a binary eutectic. T is the temperature, c the concentration of one component. 
The regions L, $\alpha$, and $\beta$ correspond to 
one-phase equilibrium states of the liquid, 
the solid $\alpha$, and the solid $\beta$ phases, respectively.
 L$+\alpha$ and L$+\beta$ are regions of two-phase equilibrium 
between the liquid and the solid phases; the actual concentrations
 of the two phases are given by the liquidus and solidus lines
 (full lines) delimiting these regions. $c_{\text{e}}$,
$c_\alpha$, and $c_\beta$
 denote the equilibrium concentrations of the liquid and
 the two solid phases at the triple or eutectic point.
 The concentrations for the undercooled case are also displayed.
} 
\vskip 0.1in
\label{fig2} 
\end{figure}

They discovered that under favourable circumstances a weak uniaxial
 stress of the order of 1 bar leads to a dramatic change in the stability range
 of the Mullins-Sekerka instability. A schematic representation of one of the 
most common liquid-solid equilibrium phase diagrams is displayed in 
Fig \ref{fig2}. Dilute alloy means that the
concentration of the minor phase is very small. The other situation,
 in which we
 are interested here, corresponds to a composition close to the eutectic
 one. The growing solid then often forms a parallel array of the two coexisting
 phases $\alpha$ and $\beta$ that grow side by side.
 This growth mode is called 
lamellar eutectic growth.

A seminal theoretical desription of lamellar eutectics has been given by  
Jackson and Hunt (JH)\cite{Jackson_Hunt66}.  Their basic idea is 
the replacement of the diffusion 
field in the liquid phase with that of a planar 
front. 
Assuming that the $\alpha$ and $\beta$ lamellas
 have equal average undercoolings
they were able to obtain an analytic 
approximation for the average undercooling 
of the interface. 
They then invoked
 the hypothesis, which has since become known as {\em minimum undercooling 
assumption}, that the {\em selected} wavelength
 of the pattern leads to the 
minimum possible value of the undercooling
 (which means that for {\em given} undercooling 
the {\em fastest-growing} structure is selected).

\section{ MODEL EQUATIONS}
In describing the problem by a macroscopic continuum model
we must introduce fields. These are the temperature,
the concentrations and  the stress fields.
We make  some standard simplifying assumptions about the
properties  of the sytem,
believed 
not to affect its essential physical features. 
These simplifications were justified elsewhere \cite{Kassner91}. 
For the sake of completeness, we recapitulate them briefly. 
The thermal gradient G is assumed 
constant in the frame of reference moving along with the growing 
interface. This means that thermal diffusion is much faster 
than chemical diffusion, that thermal conductivities of all phases 
are equal, and that latent heat production can be neglected. Thanks 
to this approximation, the motion of the temperature field is completely 
decoupled from that of the concentration field. Temperature is 
 given by position, which effectively reduces 
the number of fields to be considered by one. 
We further suppose the attachment kinetics at the solid-liquid interface 
to be fast on the time scales of all other transport processes. 
This assumption is legitimate for  microscopially rough interfaces. 
We take surface tension to be isotropic. In the vicinity of the 
operationg point in the phase diagram, the slopes of the liquidus and 
the solidus line are assumed constant. This leads to temperature independent 
partition coefficients for both phases $\alpha$ and $\beta$. The partition 
coefficients $k_{\alpha / \beta}$ are the ratios of the slopes of the 
liquidus and solidus lines, respectively. In addition, we restrict 
ourselves to the so called one-sided model, i.e., we have no diffusion 
in the solid phases.

Introducing a dimensionless concentration
 field $c=(\tilde{c}-\tilde{c}_e)/\Delta \tilde{c}$, 
where $\tilde{c}$ stands for the physical concentration and 
 $\Delta \tilde{c}=\tilde{c}_\beta-\tilde{c}_\alpha$ is the miscibility 
gap, we can write the equation of motion in the laboratory frame 
(where the sample is pushed at constant velocity V along the $-z$ direction)
\begin{equation}
\nabla^2 c + \frac{2}{l} \frac{\partial c}{\partial z} =0  \; . \label{diffuseqn}
\end{equation}
In this equation, $l=2D/V$ is the diffusion length, where
 $D$ is the diffusion constant. 
One boundary condition for the diffusion equation takes into account 
that the concentration far away from the surface has a constant value 
 $c_\infty=(\tilde{c}_\infty-\tilde{c}_e)/ \Delta \tilde{c}$.
In the
lateral direction, we assume periodic boundary
conditions:  $c(x,z)=c(x+\lambda,z)$. Mass conservation 
requires boundary conditions for the normal derivatives of the concentration 
fields at the liquid-solid interface. This continuity equation reads
\begin{equation}
-D \frac{\partial c}{\partial n} \bigg{|}_{\text{Interface}}= \left\{ 
 \begin{array}{l} 
 ((1-k_\alpha) c +\delta)   v_n    \\ 
 ((1-k_\beta)  c +\delta-1) v_n 
 \end{array}\label{continuityeqn} \right.
\end{equation}
where $\delta=(\tilde{c}_e-\tilde{c}_\alpha)/\Delta \tilde{c}$ is the reduced 
miscibility gap of the $\alpha$ phase and $1-\delta$ that of the $\beta$ phase. 
$v_n=(2D/l+\dot{\zeta}(x))n_z$ is the normal velocity of 
the interface where the normal points from the solid into the liquid. 

For the stress field we impose mechanical equilibrium, 
 $\sum_j {\partial \sigma_{ij} }/{\partial x_j} =0  $, which means
that on the time scale of the concentration field, the stress is always
relaxed. 
 We assume linear elasticity and an isotropic solid, 
so that Hooke's law reads:

\begin{equation}
 \sigma_{ij} = \frac{E_{\alpha / \beta }}{1+\nu_{\alpha / \beta }}( u_{ij} + \frac{ \nu_{\alpha / \beta }}{1-2 \nu_{\alpha / \beta }} u_{kk} \delta_{ij})   \; ,                                             
\end{equation}
where $\sigma_{ij}$ are the components of the stress tensor and 
 ${u}_{ij}=\frac{1}{2} ({\partial u_i}/{\partial x_j}+ {\partial u_j}/{\partial x_i} )$ 
those of the strain tensor ($u_i$ is the displacement vector). 
$E_\alpha$ ($E_\beta$) is Young's modulus for the $\alpha$ ($\beta$) phase,
$\nu_\alpha$ ($\nu_\beta$) the Poisson number.

The boundary conditions at the solid-liquid interface are 
\begin{eqnarray}
 \sigma_{nn}&=&{\bf n {\sigma n}}=-p_l \>, \nonumber \\
 \sigma_{nt}&=&{\bf n { \sigma t}}=0 \>,
\end{eqnarray} 
where ${\bf n}$ (${\bf t}$) is the normal 
(tangential) vector at the interface, and $p_l$ is the pressure in the liquid.
These conditions state that we have no shear at the solid-liquid 
boundary and that the normal component of the stress tensor is continuous.
That is, we neglect the capillary overpressure present when the
interface is curved. Usually, this is a good approximation.

\begin{figure}
\narrowtext
\epsfxsize=3in
\epsfbox{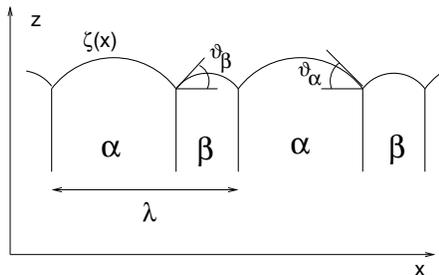}
\caption{
\noindent
Illustration of a lamellar eutectic. The interface position is
 $z=\zeta(x)$.
  The pinning angles $\vartheta_{\alpha / \beta}$ are also shown.
} 
\vskip 0.1in
\label{fig3} 
\end{figure}

There are two further points that have to be taken into account.
 Both result from the 
requirement of local thermodynamic equilibrium at the interface,
due to fast interface kinetics.
 The first of these is often referred to as `mechanical` 
equilibrium condition for the surface tensions of 
the three interfaces meeting at a triple 
point (although it is indeed a condition of thermodynamic equilibrium
under particle exchange, i.e., one of chemical equilibrium).
  The contact angles 
$\vartheta_{\alpha/\beta}$ (see Fig.\ref{fig3}) should obey  
\begin{eqnarray}
   \gamma_{\alpha l} \sin \vartheta_\alpha &+& \gamma_{\beta l} \sin \vartheta_\beta = \gamma_{\alpha \beta} \; , \label{triplepoint} \\
\gamma_{\alpha l} \cos \vartheta_\alpha &-& \gamma_{\beta l} \cos \vartheta_\beta = 0 \; , \nonumber
\end{eqnarray}
where $\gamma_{ij}$ is the surface tension between the phases $i$ and $j$
(and $l$ designates the liquid phase). 
The second condition couples the stress to the concentration field. 
It is a modified Gibbs-Thomson equation:
\begin{eqnarray}
  \epsilon_{\alpha/\beta} \; c|_{\text{interface}}  &=& \zeta / {l_T}^{\alpha /\beta} + {d_0}^{\alpha /\beta} \kappa \label{Gibbs-Thomson}\\
   &&+ \; \;{H}^{\alpha /\beta} 
 \frac{(\sigma_{tt} -\sigma_{nn})^2}{ {\sigma_0}^2} \>,  \hspace{0.5cm}\nonumber \\
 (\epsilon_\alpha=-1 \;,&& \epsilon_\beta=1) \>.\nonumber 
\end{eqnarray}
In this equation, $\zeta$ is the $z$ coordinate of the liquid-solid interface
and $\kappa$ 
its curvature, taken positive where the solid is convex. $l_T^{\alpha /\beta}$ are 
the thermal lengths,
 given by $l_T^{\alpha /\beta}=m_{\alpha /\beta} \Delta \tilde{c}/G$, 
where $m_{\alpha}$ ($m_\beta$) is the absolute value 
of the slope of the liquidus 
line describing coexistence of phase ${\alpha}$ ($\beta$) and the liquid. 
 $d_0^i=\gamma_{il} T_{\rm e}/ L_i m_i \Delta c$ are the capillary lengths $(i=
 \alpha ,\beta)$, 
 where $L_i$ is the latent heat per unit volume and $T_{\rm e}$ the eutectic 
temperature. The modification is the inclusion of the stress term with 
\begin{equation}
{H}^i= \frac{ T_{\rm e} (1-{\nu_i}^2) {\sigma_0}^2}{2 E_i |m_i| \Delta c L_i }\>; \hspace{1cm}i=\alpha,\beta \; .
\label{defHs}
\end{equation}
Herein,
 $\sigma_0$ is the uniaxial prestress that can be controlled in experiments.
A detailed derivation of  eq.~(\ref{Gibbs-Thomson}) is given in \cite{Cantat_et_al98}.

\section{Jackson-Hunt theory for a flat interface} \label{flatsurf}
\begin{figure}
\narrowtext
\epsfxsize=3in
\epsfbox{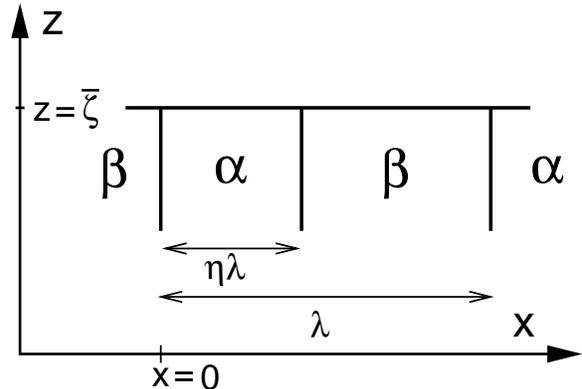}
\caption{
\noindent
The flat-interface structure used in the simplest Jackson-Hunt approach.} 
\vskip 0.1in
\label{fig3a} 
\end{figure}
The first level of approximation in Jackson and Hunt's approach consisted in 
replacing the actual diffusion field in (\ref{Gibbs-Thomson}) with that of  a
 planar lamellar structure sitting at the average position of the solidification front.
Without the stress term, (\ref{Gibbs-Thomson}) would then become a pair of second-order
differential equations with boundary conditios following from (\ref{triplepoint}).
The solution of these equations with 
the supplementary condition that the two solutions match at
the triple point gives the interface shape and the volume fraction $\eta$
of the $\alpha$ phase. Since these equations are nonlinear, they cannot easily 
be solved analytically. Hence Jackson and Hunt invoked the condition of
equal average undercooling of the two solid-liquid interfaces, which fixes
the free parameter $\eta$ and allows to obtain an analytic relation between the
average undercooling and the wavelength. The second step -- solution for the 
interface shape -- can then be done numerically, if desired.

The main modification in our case is 
that we have an additional term in (\ref{Gibbs-Thomson}) 
involving
the stress distribution at the interface. 
In the spirit of Jackson and Hunt, we compute
this expression for a flat interface first. 
Then the problem becomes very similar to
JH's original approach with the diffusion field replaced by 
$c|_i -\epsilon_{\alpha /\beta}{H}^{\alpha /\beta}
 (\sigma_{tt} -\sigma_{nn})^2/ {\sigma_0}^2 $.

Averaging the diffusion field obtained by solving the von-Neumann problem
(\ref{diffuseqn}), (\ref{continuityeqn}) for a flat interface, we have
\begin{eqnarray}
 \langle c \rangle_\alpha&=&\frac{1}{k} (c_{\infty}+\delta+\eta-1) +\frac{2 \lambda}{\eta l} P(\eta) \; , \label{uaverage1}\\
 \langle c \rangle_\beta&=&\frac{1}{k} (c_{\infty}+\delta+\eta-1) -\frac{2 \lambda}{(1-\eta) l} P(\eta) \: , \label{uaverage2}
\end{eqnarray}
where
\begin{equation}
P(\eta)=\sum_{n=1}^{\infty} \frac{ \sin^2 (n \pi \eta)}{(n \pi)^3} \; \label{defpeta}
\end{equation}
and the segregation coefficient $k$ has been taken equal in the two phases.
The averages of the curvature of the $\alpha$ and $\beta$ lamellas can be obtained
without approximation, as they just involve the integration of a derivative,
\begin{eqnarray}
 \langle \kappa \rangle_\alpha&=&\frac{2}{\eta \lambda} \sin \vartheta_\alpha \;,
\label{curvaverage1}\\
\langle \kappa \rangle_\beta&=&\frac{2}{(1-\eta) \lambda} \sin \vartheta_\beta \; .
 \label{curvaverage2}
\end{eqnarray}

To  average the stress terms, we must, in principle, solve the elastic problem
for a flat lamellar structure. Innocent as this problem may look, it is not
all that trivial. Nevertheless, the final averaging procedure will turn out to 
be independent of the subtleties that we will now 
discuss briefly. 

At each lamella boundary between the $\alpha$ and $\beta$ phases, see 
 Fig.~\ref{fig3a},
we have, on the one hand,  continuity of the normal
and shear components of the stress tensor (due to mechanical equilibrium): 
\begin{eqnarray}
\sigma_{xx}(x=0^-) &=& \sigma_{xx}(x=0^+) \; , \nonumber\\
\sigma_{xz}(x=0^-) &=& \sigma_{xz}(x=0^+) \; , \label{sigsolsol}
\end{eqnarray}
and the same conditions at $x=\eta\lambda$. On the other hand,
coherence of the interfaces between lamellas imposes additional
conditions, viz. continuity of  the displacements (up to a constant):
\begin{eqnarray}
u_x(x=0^-) &=& u_x(x=0^+) \; , \nonumber\\
u_z(x=0^-) &=& u_z(x=0^+) \; , \label{usolsol}
\end{eqnarray}
with again  identical  conditions at $x=\eta\lambda$. Equations 
(\ref{sigsolsol}) and (\ref{usolsol}) and their
counterparts at $x=\eta\lambda$ constitute two boundary conditions
at each vertical boundary
for the stress field in the lamella extending between $x=0$ and 
$x=\eta\lambda$. (There are four equations but each of them 
pertains to two lamellas.) The four boundary conditions at
the two $x=\hbox{const.}$ boundaries of a lamella 
 suffice to solve the elastic problem uniquely. Therefore, there is no room
left for more boundary conditions. But in fact, we have, at the
boundary towards the liquid
\begin{eqnarray}
\sigma_{zz}(z=\bar\zeta) &=& - p_l \; , \nonumber\\
\sigma_{xz}(z=\bar\zeta) &=& 0  \; , \label{sigliqsol}
\end{eqnarray}
two additional boundary conditions, rendering the problem overdetermined.
Note that this line of reasoning presupposes different elastic constants
in the solid phases. If all elastic coefficients are equal, then
the validity of  (\ref{usolsol})  implies that of  (\ref{sigsolsol})
simply by virtue of Hooke's law
 (assuming, as usual, that continuous physical
functions are  also continuously differentiable). 
With different sets of elastic constants
in the two phases, we have a situation similar to that in microstructures
 discussed
by M\"uller \cite{mueller98}. A solution to the elastic problem need not exist.
That is, the elastic problem may not have a solution with the
boundaries {\em fixed} to the chosen positions. However, 
a solution to the mathematical problem given all the discussed
boundary conditions does exist, if we allow
the lamella boundaries to adjust their shape, i.e., if we convert the
question to a free-boundary problem. The purpose of the following
discussion is then only to establish that analytically tractable
{\em homogeneous-stress solutions} exist in {\em particular} cases.  

In fact, we do not need {\em general}
 solvability to consider a sensible
physical problem. Looking for constant-stress solutions of (\ref{usolsol})
together with (\ref{sigliqsol})
we obtain, setting $\sigma_{xz}(\bar{\zeta})=0$, the 
conditions
\begin{eqnarray}
-p_l=\sigma_{zz}^{0} &=& \frac{\nu_\alpha}{1-\nu_\alpha} \sigma_{xx}^0
+ \frac{E_\alpha}{1-\nu_\alpha^2} u_{zz}^0 \; , \nonumber\\
-p_l=\sigma_{zz}^{0} &=& \frac{\nu_\beta}{1-\nu_\beta} \sigma_{xx}^0
+ \frac{E_\beta}{1-\nu_\beta^2} u_{zz}^0 \; , \label{condsol}
\end{eqnarray}
where the superscript 0 indicates the absence of spatial variation
inside the lamellas and the
subscripts $\alpha$ and $\beta$ distinguish the elastic constants in 
the two solid phases. There are no such subscripts on the stresses
and on $u_{zz}^0$ which are equal in the two phases (in contrast to $u_{xx}^0$,
which may differ). It is evident that for different elastic constants
in the two materials, (\ref{condsol}) has a unique solution for $\sigma_{xx}^0$
and $u_{zz}^0$, providing the coefficient determinant does not vanish.
That is, we just have to choose the right value of the prestress 
$\sigma_{xx}^0$ to ensure the existence of a homogeneous
solution on which we can
base our analysis \cite{footnote1}.
 As long as $p_l\ne0$, we have $\sigma_{xx}^0\ne -p_l$,
i.e. the Grinfeld instability is potentially activated. For $p_l = 0$,
on the other hand, we can even have a continous set of solutions, if we
choose the elastic constants such that the coefficient determinant 
vanishes (which is possible even for $E_\alpha\ne E_\beta$, say).

Given the fact that there is a solution to the elastic problem, the calculation
of its influence on the Gibbs-Thomson equation  (\ref{Gibbs-Thomson})
becomes very simple. As $\sigma_{xx}$ is homogeneous throughout
the sample and because of $\sigma_{tt}=\sigma_{xx}$ for a planar interface,
we simply have $(\sigma_{tt}-\sigma_{nn})^2/\sigma_0^2=1$. Hence, the
averaged stress terms are simply ${H}^{\alpha}$ and $H^\beta$, respectively.
 
Inserting this in the Gibbs-Thomson equation, we get
\begin{eqnarray}
 \langle \zeta \rangle_\alpha &=& {\langle \zeta \rangle_\alpha}^{JH}+{l_T}^\alpha {H}^{\alpha} \; , \nonumber\\
\langle \zeta \rangle_\beta &=& {\langle \zeta \rangle_\beta}^{JH}+{l_T}^\beta {H}^{\beta}  \; , \label{modifav}
\end{eqnarray}
where ${\langle \; \; \rangle}^{JH}$ is the average without 
the stress term.
Assuming equal average undercoolings in front of both phases, we
  set $\langle \zeta \rangle_\alpha=\langle \zeta \rangle_\beta$,
 (because $\Delta T = -G \zeta$). As has been discussed earlier,
this assumption 
is not necessary to obtain closed equations
\cite{Kassner91}, but it simplifies calculations. 
We can then write an implicit equation for $\eta$:
\begin{eqnarray}
 \eta&=&1-c_\infty-\delta+k \frac{ {l_T}^\beta {H}^{\beta} -{l_T}^\alpha {H}^{\alpha}}{{l_T}^\alpha + {l_T}^\beta }  \nonumber \\ 
 &+&  \frac{k}{ ({l_T}^\alpha + {l_T}^\beta) \; \eta \; (1-\eta)} 
 \{ \frac{2 \lambda}{l} P(\eta) [ \eta {l_T}^\beta -(1-\eta) {l_T}^\alpha] \nonumber \\ &&+ \frac{2}{\lambda} [ \eta {l_T}^\beta {d_0}^\beta \sin \vartheta_\beta- (1-\eta) {l_T}^\alpha  \sin \vartheta_\alpha ] \} \;
 .\label{flateta}
\end{eqnarray}
 The last term in this equation  is small
 for small undercooling (implying small P\'eclet number $\lambda / l$)
 and small contact angles, so that in this limit an explicit formula for 
$\eta$ is available.
 Using (\ref{flateta}) in (\ref{modifav}),
 we obtain for the averaged undercooling:
\begin{equation}
\langle \Delta T (\lambda) \rangle=\langle \Delta T \rangle_{\text{min}} 
 \left( \frac{\lambda}{\lambda_{\text{min}}}+\frac{\lambda_{\text{min}}}{\lambda} 
\right) \; ,
\end{equation}
where 
\begin{eqnarray}
 \lambda_{\text{min}}&=&{\lambda_{\text{min}}}^{JH}(\eta)  \label{lambdamin}\; ,\\
 \langle \Delta T \rangle_{\text{min}}
 &=&{\langle \Delta T \rangle_{\text{min}}}^{JH}
+G \frac{{l_T}^\alpha  {l_T}^\beta}{{l_T}^\alpha + {l_T}^\beta}
 ({H}^{\alpha} +{H}^{\beta}) \; .
\end{eqnarray}
Because on setting $\partial \langle\Delta T \rangle/ \partial \lambda =0$ the elastic
terms disappear from the equation for $\lambda_{\text{min}}$, there seems 
at first glance 
to be no effect of elasticity on the selected wavelength.
 But that is not true, because
 $\eta$ has changed. 
  Expanding ${\lambda_{\text{min}}}$ about $\eta^{JH}$, setting 
 $\eta=\eta^{JH}+\Delta \eta$, we obtain
\begin{eqnarray}
{\lambda_{\text{min}}}&=&{\lambda_{\text{min}}}^{JH}(\eta^{JH})  \label{flatlambda}
\biggl( 1+\Delta \eta \; [ - \frac{1}{2} \frac{ P'(\eta^{JH})}{P(\eta^{JH})} \nonumber \\&&+ \frac{ {d_0}^\beta \sin \vartheta_\beta - {d_0}^\alpha \sin \vartheta_\alpha}{ \eta^{JH}  {d_0}^\beta \sin \vartheta_\beta + (1- \eta^{JH}) {d_0}^\alpha \sin \vartheta_\alpha} ] \; \biggr) \; , \label{lamflat}
\end{eqnarray}
where
 $\Delta\eta\approx k ({l_T}^\beta {H}^{\beta} -{l_T}^\alpha {H}^{\alpha})/({l_T}^\alpha + {l_T}^\beta)$. 

  The first thing to note is that if the elastic constants and the latent heat per volume are equal 
 in the two phases,
elastic effects do not influence the wavelength at minimum undercooling,
within the flat-interface approximation. This is why we insisted on considering
the more general case in spite of the complications concerning the existence
of a solution to the elastic problem. The logarithmic derivative 
$P'(\eta)/P(\eta)$ of the JH function 
 vanishes for $\eta=\frac{1}{2}$ and diverges for $\eta\to0$ or 
$\eta\to1$, allowing for a potentially large effect. However, it stays 
smaller than 50 for $0.04 < \eta < 0.96$, which means that it does not
provide more than an order of magnitude in most situations. The second 
term in the brackets of (\ref{lamflat}) usually is on the order of one.
The sign of the effect depends on the sign of $\Delta \eta$, i.e, the 
relative magnitude of the elastic constants in the two phases.

If we assume that  the difference in Young's moduli in the two phases 
is on the order of 
10\% of their average (i.e., 
$(1-\nu_\beta^2)/2E_\beta-(1-\nu_\alpha^2)/2E_\alpha\approx 0.05/E_{\text{av}} $),
we find, for typical values
of the material parameters ($T_{\rm e}\sim 400$K, a freezing range
 $m_i\Delta c\sim10K$, $L_i\sim10$J/cm$^3$, $k\sim1$, $E={10}^5$ N/{cm}$^2$)
and for $\eta\sim 0.1$
that 
 $\Delta\eta P'(\eta)/P(\eta) \approx 2\times 10^{-7}$ [cm$^4$/N$^2$] $\sigma_0^2$.
This gives a relative wavelength change of 
10$^{-5}$ for 
 $\sigma_0=1$ bar and one of 
 10\% for $\sigma_0=100$ bar. We therefore 
conclude that this effect is small in ordinary experiments but might
be accessible in high-pressure setups, where pressures of 
100 bar or more could be applied.

The next task is then to see what is the order of 
magnitude of  the influence of deviations of
the interface shape from planarity.

 \section{Jackson-Hunt theory for a triangular interface}
\begin{figure}
\narrowtext
\epsfxsize=3in
\epsfbox{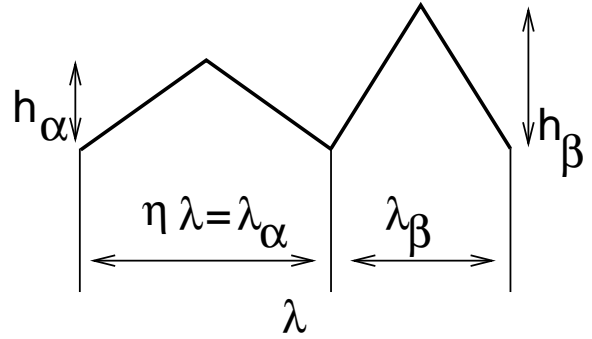}
\caption{
\noindent
Simplified surface structure} 
\vskip 0.1in
\label{fig4} 
\end{figure}
 The simplest non-planar surface structure  accessible to an analytic
 approach 
is a triangular surface (see Fig \ref{fig4}).
To proceed, we will from now on assume
that the elastic constants are the same in the two phases.

In the absence of volume forces, 
 the two-dimensional stress tensor
can be expressed via an Airy stress function $\chi$.
Setting
\begin{equation}
  \sigma_{xx}= \frac{\partial^2 \chi}{\partial z^2} \; ,\hspace{0.5cm} 
  \sigma_{xz}=-\frac{\partial^2 \chi}{\partial z \partial x} \; , \hspace{0.5cm} 
  \sigma_{zz}= \frac{\partial^2 \chi}{\partial x^2}  \; , 
\end{equation}
we automatically satisfy the condition of 
mechanical equilibrium $\sum_j \partial\sigma_{ij}/\partial x_j =0$.
Hooke's law together with the assumption of isotropic elastic properties
then implies that $\chi$ must obey the
biharmonic equation $\Delta^2 \chi=0$.
We split the Airy function according to $\chi(x,z)= \chi^{(0)}(x,z) + \chi^{(1)}(x,z)$, where 
 \begin{eqnarray}
 \chi^{(0)}(x,z) &=& -\frac{p_l}{2} x^2 + \frac{\sigma_0-p_l}{2} z^2 \;\label{chi0} , \\
 \chi^{(1)}(x,z) &=&  \sum_{n=1}^\infty  ( {A_n} z +B_n) e^{K_n z} e^{i K_n x} + c.c. \label{chi1} \;, 
\end{eqnarray} 
 $ K_n={2 \pi n}/{\lambda}$, and both terms are solutions to the biharmonic 
equation separately.

Equation (\ref{chi0}) corresponds to a homogeneous stress 
state and (\ref{chi1}) describes 
the deviation therefrom.
 Once we have calculated the coefficients $A_n$, $B_n$ we 
are able to compute the stress term in (\ref{Gibbs-Thomson}).
 Inserting our boundary conditions for the 
 stress field into a representation of $\sigma_{nn}$ and $\sigma_{nt}$ 
 in the $xz$ coordinate sytem, we arrive at an infinite 
 linear system of equations that in principle could be solved 
 for the coefficients. 
 An analytic result can be obtained, if the equations are expanded in 
 terms of 
 $\Delta_{\alpha / \beta}=2 h_{\alpha / \beta} / \lambda_{\alpha / \beta}$, 
where $\lambda_\alpha=\eta\lambda$ and
 $\lambda_\beta=(1-\eta)\lambda$
 are the widths of the lamellas and 
 $h_{\alpha}$ ($h_{\beta}$) is the height of the triangle in the
 $\alpha$ ($\beta$) phase (Fig.~\ref{fig4}).
If the expansion is performed up to linear order, one arrives at
\begin{equation}
 {A_n}= -\sigma_0  \lambda e^{-i \pi n \eta} {\Delta_{\alpha \beta}}^n (\eta) \; , \hspace{0.5cm} {B_n}=0 \; , \label{an}
\end{equation}
where
\begin{eqnarray}
 {\Delta_{\alpha \beta}}^n (\eta)&=&
 \delta_{n,0} \left(-\Delta_\alpha\frac{\eta}{2} 
            +\frac{1}{4}\Delta_\alpha \eta^2 
            +\frac{1}{4}\Delta_\beta (1-\eta)^2\right) \nonumber \\
 &&+(1- \delta_{n,0}) \; \;\frac{1}{2 \pi^2 n^2} \nonumber \\
 &&        \left[ \Delta_\alpha+\Delta_\beta (-1)^n
             -(\Delta_\alpha+\Delta_\beta) \cos(\pi \eta n)\right].
\end{eqnarray}
Note that in (\ref{an}) we need this definition only for $n>0$,  
where it simplifies to the second term.
Using these coefficients in $\chi$ and calculating
 the average of $(\sigma_{tt} -\sigma_{nn})^2$, 
we obtain

\begin{eqnarray}
\langle (\sigma_{tt} -\sigma_{nn})^2 \rangle_\alpha &=&{\sigma_0}^2 \left[1-\frac{1}{\eta} \Omega (\eta) \right]         \; ,  \\
\langle (\sigma_{tt} -\sigma_{nn})^2 \rangle_\beta &=&{\sigma_0}^2 \left[1+\frac{1}{(1-\eta) } \Omega (\eta)  \right]       \; ,
\end{eqnarray}
where
\begin{equation} 
\Omega (\eta)= 16 \sum_{n=1}^{\infty} \sin(\pi n \eta)  {\Delta_{\alpha \beta}}^n (\eta)\>.
\end{equation}

To be consistent,
 we have to compute the average of the diffusion field for the 
double trianglular surface as well. It turns out that the result can be
cast into a form that is very similar to the case of a planar interface. 
All that has to be done is to replace the Jackson-Hunt function $P(\eta)$
by
\begin{eqnarray}
P(\eta,\Delta_\alpha,\Delta_\beta)
&=&P(\eta)  + \frac{2}{\pi^2} \sum_{n=1}^{\infty} \frac{\sin \pi \eta n}{n} \nonumber \\
 &&\sum_{m=1}^{\infty} \frac{\sin \pi \eta m}{m}
 ( \Delta_{\alpha\beta}^{n-m}(\eta)-\Delta_{\alpha\beta}^{n+m}(\eta) )
\; ,\label{defpmodeta}
\end{eqnarray}
and here all integer values, including zero, can appear in the superscript of
 $\Delta_{\alpha\beta}^{n-m}(\eta)$. Whereas $P(\eta)$ is essentially
independent of $\lambda$, the wavelength dependence of $\eta$ being 
weak, $P(\eta,\Delta_\alpha,\Delta_\beta)$ does depend on the wavelength 
via the $\lambda$ dependence of the $\Delta_{\alpha/\beta}$.
This must be taken into account in the minimization procedure when the
minimum undercooling is determined.

Thus replacing $P(\eta)$ with $P(\eta,\Delta_\alpha,\Delta_\beta)$
in (\ref{uaverage1}) and (\ref{uaverage2}), we can proceed in a pretty
straightforward manner. First we use an assumption analogous to
the equal undercooling assumption to eliminate the term
 $ \frac{1}{k} (c_{\infty}+\delta+\eta-1)$ from the formulas.
In particular, we assume 
 $ \langle \zeta \rangle_\alpha - \langle \zeta \rangle_\beta = \frac{1}{2} (h_\alpha -h_\beta) $. 
Next, we write down the total average undercooling.
In minimizing it, we suppose a weak $\lambda$ dependence of $\eta$,
which 
yields $\partial P(\eta,\Delta_\alpha,\Delta_\beta)/\partial \lambda
= -P_1(\eta,\Delta_\alpha,\Delta_\beta)/\lambda$ with
 $P_1(\eta,\Delta_\alpha,\Delta_\beta)\equiv  P(\eta,\Delta_\alpha,\Delta_\beta)-P(\eta)$.
We then find that, surprisingly, the result for the wavelength 
does not contain the modified Jackson-Hunt function anymore but just 
the original one:
\begin{eqnarray}
{\lambda_{\rm min}}^2 =&& \frac{l}{P(\eta)} \{ d_0^\alpha (1-\eta)
 \sin \vartheta_\alpha+d_0^\beta \eta \sin \vartheta_\beta  \nonumber \\
&+& \frac{1}{2}(\eta H^\beta- (1-\eta) H^\alpha) \tilde\Omega(\eta) \}
\label{wavelengthnew} \; ,
\end{eqnarray}
where 
\begin{eqnarray}
\tilde\Omega(\eta)&&\equiv \lambda\Omega(\eta) =  8\sum_{n=1}^\infty
\frac{\sin( \pi \eta n) }{ \pi^2 n^2} \nonumber \\
       && \left[\frac{h_\alpha}{\eta}+\frac{h_\beta}{1-\eta} (-1)^n
             -\left(\frac{h_\alpha}{\eta}+\frac{h_\beta}{1-\eta}\right) \cos(\pi \eta n)\right]\>.
\label{defomtilde}
\end{eqnarray}
For comparison  with 
the stress-free case we rewrite this as
\begin{equation}
{\lambda_{\rm min}}^2={ {\lambda}^{JH}_{\rm min}}^2(\eta) \left( 1+ \frac{(\eta H^\beta- (1-\eta) H^\alpha) \tilde\Omega(\eta)}{2[
{d_0}^\alpha (1-\eta) \sin \vartheta_\alpha+{d_0}^\beta \eta \sin \vartheta_\beta]}\right) \; , \label{wavelenminJH}
\end{equation}
where we have taken the Jackson-Hunt result for the wavelength at the
pertinent value of $\eta$. Of course, there is an additional effect
(as in Sec.~III) due to the change in the 
 volume fraction under external stress. The latter is 
given by
\begin{eqnarray}
 \Delta \eta= &\;& \frac{k}{{l_T}^\alpha+{l_T}^\beta } \Biggl\{ 
 ({l_T}^\beta H^\beta -{l_T}^\alpha H^\alpha) +\frac{1}{2} (h_\beta-h_\alpha) \nonumber \\
   &+& \left(\frac{{l_T}^\beta}{1-\eta}-\frac{{l_T}^\alpha}{\eta}\right)
       \frac{2\lambda}{l} P_1(\eta,\Delta_\alpha,\Delta_\beta) \label{etachange} \\
  &+&\Omega(\eta) \left(\frac{{l_T}^\beta}{1-\eta} H^\beta +\frac{{l_T}^\alpha}{\eta} H^\alpha \right) \Biggr\} \nonumber \; .
\end{eqnarray}

In order to get an estimate of the
order of magnitude
 of elastic effects, we note that for $\sigma_0\approx 1$~bar and the 
material parameters considered in section III,
we have $H^{\alpha/\beta}\approx2\times 10^{-5}$. $\Omega(\eta)$ is on 
the order of  ten, hence $\tilde\Omega(\eta)\approx\lambda$, if we
take the heights $h^{\alpha/\beta}$ of the lamellas
 to be of order $\lambda/10$.
 Assuming $d_0^{\alpha/\beta}\approx 10^{-3}\lambda$,
we find that the second term in (\ref{wavelenminJH}) is on the order
of one percent for $\sigma_0=1$ bar, i.e., an appreciable effect may
be expected for pressures or tensions in excess of 10 bar.

With the same assumptions, we note that 
the change of  $\eta$ induced by elastic effects is on the order of 
 $10^{-4}$ for $\sigma_0=1$~bar and $10^{-2}$ for $\sigma_0=10$~bar,
hence negligible in most cases in comparison with the direct effect
given by (\ref{wavelenminJH}). Of course, this also depends on the
size of $d \lambda_{\rm min}^{JH}/d\eta$, which we have estimated to be 
small for $\eta$ values not too close to 0 or 1, in Sec.~III.

We now consider a few special cases that are especially transparent.

If the lamella structure
 is symmetric under an exchange of the $\alpha$ and $\beta$ phases,
i.e., $\eta=\frac{1}{2}$ and $h_\alpha=h_\beta$, then we see immediately
from (\ref{defomtilde}) that $\tilde\Omega(\eta)=0$. Terms with even $n$
vanish because of the factor $\sin(\pi\eta n)$, terms with odd $n$
produce a factor
of zero inside the brackets. Therefore, application of external stress
will not alter the wavelength in this case, except possibly via the 
change in $\eta$ induced by (\ref{etachange}),
which is a much smaller effect. Moreover, if we assume
the {\em thermal} properties of the two phases to be the same, i.e.
 $l_T^\alpha= l_T^\beta$, $L_\alpha=L_\beta$, we have $H^\alpha=H^\beta$
according to (\ref{defHs}) (because we took the {\em elastic} properties of 
both phases equal from the outset of this section). Therefore, we have
 $\Delta\eta=0$ in this case. The direct effect on $\lambda$ as described
by (\ref{wavelenminJH}) is then absent even if $h_\alpha\ne h_\beta$,
although there will be a small shift in $\eta$, if the two phases have
different heights. 

Another simplification arises, if we choose all the properties of
the $\alpha$ and $\beta$ phases to be equal and set
 $\Delta_\alpha=\Delta_\beta\equiv\Delta$ but allow for $\eta\ne\frac{1}{2}$.
In particular, this means that we assume the heights of the lamellas
to be proportional to their widths.  We can then evaluate 
 $\Omega(\eta)$ analytically,
\begin{eqnarray}
\Omega(\eta)&= &  \frac{8\Delta}{\pi^2} \sum_{n=1}^{\infty} 
\frac{\sin(\pi n \eta)}{n^2} \Bigl[1+(-1)^n-2 \cos(\pi n \eta)\Bigr] \nonumber \\
&=& \frac{8\Delta}{\pi} \left( \eta \ln 2+ \int_{0}^{\eta} dx \ln |\sin( \pi x)|\right)\;,
\label{omegadd}
\end{eqnarray}
and it is easy to show that $(2\eta-1)\Omega(\eta) \ge 0$.
Therefore, we have an {\em increase} of the wavelength in this case. 

\begin{figure}
\narrowtext
\epsfxsize=3in
\epsfbox{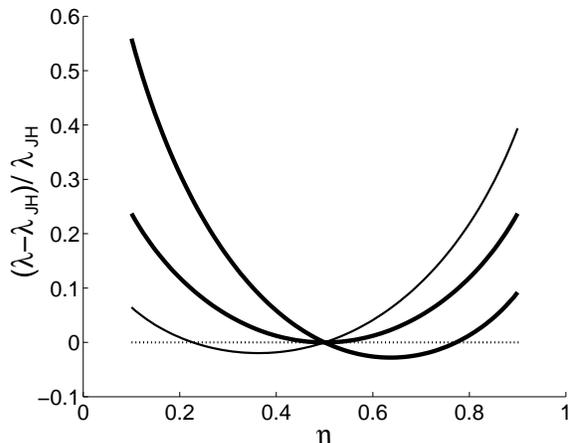}
\caption{\noindent
The change in wavelength $\lambda$ as a function of the volume fraction $\eta$ for $25\;bar$.
The thick symmetric curve is for $\Delta_\alpha=\Delta_\beta=0.1$. The thick asymmetric curve is for $\Delta_\alpha=0.1$ and $\Delta_\beta=0.2$ and the thin curve is for $\Delta_\alpha=0.1$ and $\Delta_\beta=0.05$. $\vartheta_{\alpha /\beta} = \arctan \Delta_{\alpha / \beta}$ is assumed.  }
\label{lambdadeviation2}
\end{figure}

A discussion of the general case is most easily done by numerical evaluation of
(\ref{wavelengthnew}) for a few characteristic sets of parameter values
and graphical representation of the result.
 This is carried out in Fig.~\ref{lambdadeviation2}. 
 We compare the $\eta$ dependence of 
the relative change in wavelength for
 $\Delta_\alpha=\Delta_\beta$, $\Delta_\alpha=2 \Delta_\beta$ and 
 $\Delta_\beta=2 \Delta_\alpha$.
  $\Delta_\alpha$ is set to 1/10 and the pressure is $25$~bar. 
The diffusion length is taken to be $l =10^2 \lambda$ and the capillary length 
 $d_0={10}^{-3} \lambda$. 
The contact angles have been chosen as 
 $\vartheta_{\alpha /\beta} = \arctan \Delta_{\alpha / \beta}$ in
keeping with the spirit of the triangular approximation.  
 It is seen that when there is an asymmetry between the lamellas,
 a decrease of the wavelength can occur, but the magnitude 
of the effect is pretty small if $\Delta_\alpha \approx \Delta_\beta$.       
\section{Summary}
To conclude, motivated by the 
fact that the interaction between the Grinfeld and Mullins-Sekerka
instabilities is strong in directional solidification of dilute alloys
\cite{Durand96,Cantat_et_al98}, we were led to investigate the influence
of uniaxial stress in directional solidification of lamellar eutectics. 

From the outset, two differences could be expected. First, the basic
lamellar structure is not determined by the MS instability,
so direct visibility of an  interaction with the ATG instability was not
likely. Second, since the lamellar spacing is typically an order of 
magnitude smaller than cell spacings in dilute alloys, the influence
of the ATG instability which at typical thermal gradients is ``resonant''
with the MS instability should be expected to be weaker in eutectics.

On the other hand, it is also known that qualitative features that are
present in dilute alloys, such as parity breaking or the appearance
of asymmetric cells, invariably turn up in eutectics, too, albeit often
via a different mechanism, which is a rather fascinating phenomenon by itself. 
Parity breaking, for example, can be explained by
two-mode coupling in cellular growth but requires  quite a different 
analytic approach in the case of eutectics  \cite{valance93}.
More basic features, such as the underlying symmetries, are the same
in the two cases.

A similar situation arises here: The mechanism, by which stress modifies
the properties of
 the system is entirely different from that of the dilute-alloy case.
There it was the coupling to the MS instability, here it is a 
coupling to {\em the asymmetry between the two solid phases}.
Uniaxial stress has a direct effect on the volume fraction of the
phases, which in general results in a (small) influence on the wavelength
of the pattern. In addition, it changes the undercooling of the front in a 
wavelength-dependent manner, provided there is a (geometric) 
difference between the $\alpha$ and $\beta$ phases. Both effects
were calculated to linear order in the deviation $\Delta$ of the
front shape from planarity. The first effect is present even for
a planar interface, if the elastic constants of the two solid phases
differ, and it has been evaluated for that case as well.

As expected, appreciable wavelength changes require stresses that exceed
those necessary in dilute alloys by an order of magnitude. So we do not
expect elastic effects to strongly affect directional solidification
 experiments with eutectics  
 by accident (which might however happen for dilute-alloy experiments).
Nevertheless, stresses of 25 bar or so are not too high to be imposed in 
a controlled experiment which then would allow to test this theory. 
 
Another point worth mentioning is that the wavelength change can be
both positive and negative for eutectics (and is positive most of the
time) whereas we  have only seen a wavelength decrease with dilute
alloys so far (at small pulling velocities, the case considered here).
This makes the effect somewhat less interesting for material processing
purposes but underlines the basic difference in the mechanisms by 
which stress modifies microstructures in the two cases.
Large stresses ($>$ 100 bar), however, might be used to engineer the
volume fraction of the phases -- if they can be sustained in an appropriate
experimental setup.

This work was supported by a 'PROCOPE' grant in the framework of a
 French-German cooperation.



\begin{references}
\bibitem{footnote1} What happens physically, 
if we try to impose an external stress
in the $x$ direction that is different from the one giving rise to a 
homogeneous solution, is an altogether different matter.
Normally, we then have to deal with the aforementioned free-boundary problem.
Physically, it is of course also possible that
the coherence of the lamellar interfaces
gets lost, i.e., the solid phases may slip on each other, which 
may strongly modify the local crystal structure.
\end{references}
\end{document}